# Denoising Scanning Tunneling Microscopy Images of Graphene with Supervised Machine Learning

Frédéric Joucken[1], John L. Davenport[2], Zhehao Ge[2], Eberth A. Quezada-Lopez[2], Takashi Taniguchi[3], Kenji Watanabe[4], Jairo Velasco Jr.[2], Jérôme Lagoute[5], and Robert A. Kaindl[1]

[1]*Department of Physics, Arizona State University, Tempe, AZ 85287, USA*

[2]*Department of Physics, University of California, Santa Cruz, CA 95064, USA*

[3]*International Center for Materials Nanoarchitectonics, National Institute for Materials Science, 1-1 Namiki, Tsukuba 305-0044, Japan*

[4]*Research Center for Functional Materials, National Institute for Materials Science, 1-1 Namiki, Tsukuba 305-0044, Japan*

[5]*Laboratoire Matériaux et Phénomènes Quantiques, UMR 7162, Université Paris Diderot – Paris 7, Sorbonne Paris Cité, CNRS, UMR 7162 case courrier 7021, 75205 Paris 13, France*

**Abstract.** Machine learning (ML) methods are extraordinarily successful at denoising photographic images. The application of such denoising methods to scientific images is, however, often complicated by the difficulty in experimentally obtaining a suitable expected result as an input to training the ML network. Here, we propose and demonstrate a simulation-based approach to address this challenge for denoising atomic-scale scanning tunneling microscopy (STM) images, which consists of training a convolutional neural network on STM images simulated based on a tight-binding electronic structure model. As model materials, we consider graphite and its mono- and few-layer counterpart, graphene. With the goal of applying it to any experimental STM image obtained on graphitic systems, the network was trained on a set of simulated images with varying characteristics such as tip height, sample bias, atomic-scale defects, and non-linear background. Denoising of both simulated and experimental images with this approach is compared to that of commonly-used filters, revealing a superior outcome of the ML method in the removal of noise as well as scanning artifacts – including on features not simulated in the training set. An extension to larger STM images is further discussed, along with intrinsic limitations arising from training set biases that discourage application to fundamentally unknown surface features. The approach demonstrated here provides an effective way to remove noise and artifacts from typical STM images, yielding the basis for further feature discernment and automated processing.

## I. **INTRODUCTION**

Since its first demonstration by Binnig and Rohrer [1], scanning tunneling microscopy (STM) has become a powerful and essential tool in materials science due to its ability to measure and image electronic structure with sub-angstrom resolution [2]. Although the present paper focuses on imaging, STM is used for many other purposes, including atomic-scale electronic spectroscopy, atom manipulation, or determination of momentum-resolved electronic band structure [3–5]. The recent rise of machine learning (ML) has spawned completely new approaches for analyzing complex scientific data [6–9]. First applications of ML to STM-related problems have also been reported, enabling automated identification and classification of atomic-scale patterns, materials phases, and defects [10–14] as well as advances towards autonomous STM operation through ML-based tip shaping and lithography [13,15,16].

The problem of reducing noise in STM images – including both statistical noise and artifacts - has, however, not yet been addressed. In the processing of STM images for analysis, the reduction of such noise provides several benefits. First, the enhancement of the signal-to-noise ratio can improve the level of confidence when interpreting experimental STM data. Secondly, STM measurements are inherently slow due to their scanning nature, and many acquired images are discarded by the experimentalists because of limited quality achieved. Thus, an efficient ML-based filter could increase the yield of STM experiments and the amount of data available for analysis. Finally, it can be expected that denoising will enhance the performance of other subsequent computational or ML-based steps such as defect recognition, autonomous data acquisition, and control.

Supervised ML models based on convolutional neural networks (CNN) have been incredibly successful at denoising photographic images [17–27]. The common strategy for achieving high performance denoising of photographic images is to train the network on images onto which noise (typically Gaussian) is added intentionally. The corrupted images are then used as the inputs of the CNN while the original images are used as the expected result, also referred to as "ground-truth" (GT) in the models. However, this strategy cannot be implemented straightforwardly for denoising scientific images because the expected result of scientific images is more difficult to assess. For some experimental methods, it is possible to produce such fundamental images by simply acquiring images for a long time [28,29]. This assumes that both statistical noise and artifacts diminish with increased scanning. However, in STM imaging even images with low statistical noise can be far from the expected GT representation because of the role played by the tip in the imaging process. Indeed, the density of states of the tip is convolved with the density of states of the sample to produce the image [30]. Even if efforts are generally put in acquiring images with tips that have a constant or known density of states so that their effect on the imaging is understood [31–35], it is always difficult to assess whether an STM image can be considered as the true representation of the underlying electronic structure.

As a guide, experimental STM results are often compared to simulated ones to assess their validity. The simplest way to simulate an STM image is to completely neglect the effect of the tip and assume that the image is proportional to the local density of states (LDOS) of the sample [30]. Here, we exploit this idea by training a CNN on intentionally corrupted images simulated using the tight-binding approach for computing the LDOS. The tight-binding method allows for producing a large training set in a reasonable amount of time, contrary to more computationally demanding electronic structure computation techniques such as density functional theory. The trained CNN is then used to denoise experimental STM images, as illustrated in Fig. 1(a). We compare this approach to other denoising techniques, and we discuss the advantages and inconveniences offered by our simulation-based denoising technique for STM.

Simulation-based denoising was recently introduced by Mohan *et al*. for the case of transmission electron microscopy (TEM) images [36]. However, both the nature of the electronic interaction underlying the TEM imaging process as well as the generation of the training set images differs fundamentally from that required for STM denoising. Moreover, we aim at training a model that can be applied to STM images of an entire family of materials encompassing both graphite and its two-dimensional counterparts of mono- and multi-layer graphene. As discussed below, this requires the generation of a very diverse training set.

Another approach which has naturally been popular with scientific denoising is unsupervised learning, where the need for a GT image as the expected result is removed altogether [37–42]. It has been successfully applied to fluorescence microscopy [37–39,42] and TEM data [41]. Although unsupervised approaches are appealing for denoising STM images, they are challenging to adapt to STM for two reasons. First, obtaining a large training set is difficult in STM because the time needed to acquire an image is relatively long compared to other microscopy techniques. Secondly, some assumptions on the nature of the noise must be made to build the loss function in unsupervised learning [40]. Although some of the noise in STM images can be assumed to take a certain statistical form [43], the influence of the tip shape and density of states cannot be accounted for by assuming statistical noise.

## II. COMPUTATIONAL MODEL

For the CNN, the model is implemented in Python with the Keras [44] software library and its TensorFlow backend [45]. The architecture is built on the U-Net type of networks [46]. which was shown by Mohan *et al.* to be most efficient for denoising corrupted simulated images [36]. Our adapted version of the U-Net architecture is depicted in Fig. 1(b). It consists of two 'down' building blocks between which the size of the arrays is divided by two by the Keras `maxpooling` operation, a lower building block, and two 'up' building blocks between which the size of the arrays is multiplied by two by the `upsampling` operation. Each block has two convolution layers with $x_F$ filters rectified linear unit (`ReLU`) activation functions that introduce the non-linearity of the network, followed by `batchnormalization` which centers the mean output of each block close to 0 and its standard deviation close to 1. The size of the 2D convolution kernel is set to $3 \times 3$ and the number of filters in the first block of each layer was $n$ = 32 and is doubled or halved, respectively, after each `maxpooling` or `upsampling` operation. The last layer has a `tanh` activation function to best adapt the overall output range to that of the GT images (so-called labels). As the loss function, i.e. quantity that is minimized during training, we choose the mean absolute error between the outputs and labels. To optimize the network, we used the `Adam` algorithm in Keras, which considers the exponentially-weighted average of the loss function gradient [47]. The learning rate that sets the step size of the gradient descent algorithm was set to 0.001 per iteration. A typical training run was done on $3\times10^4$ simulated images with an additional 500 images used for validation at each iteration. Figure 1(c) shows the evolution of the loss function with the iteration (epoch) number during training, along with the loss computed on the validation images.

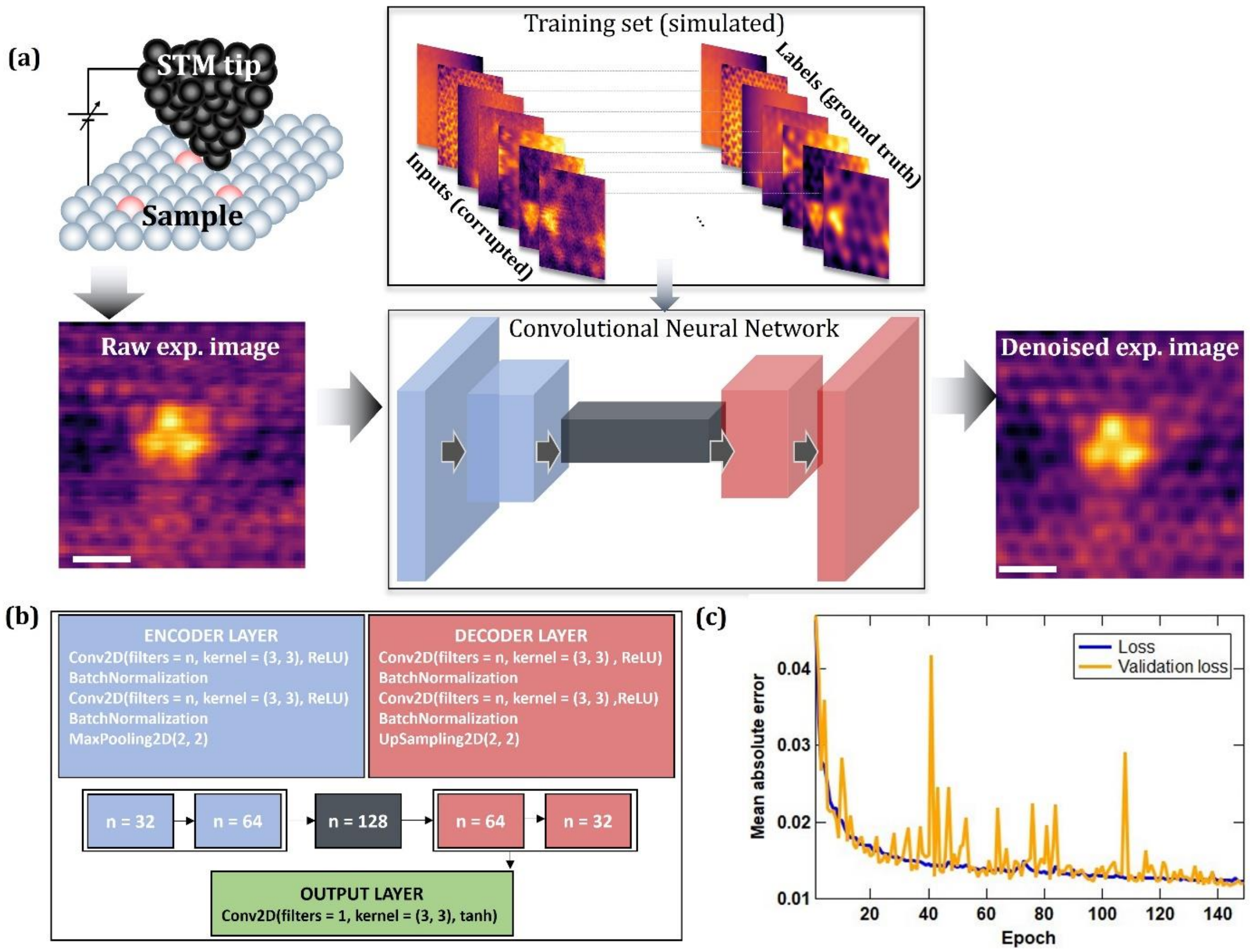

**Figure 1. Overview of our machine learning (ML) denoising approach applied to STM images. (a)** Workflow of the ML denoising based on simulated STM images. The convolutional neural network (CNN) is trained on a set of simulated STM images which are intentionally corrupted, and then applied to experimental images. Scale bars of 0.5 nm. **(b)** Architecture of the CNN, which includes two encoding layers, one bottom layer, and two decoding layers with details given in the text. **(c)** Evolution of the loss function i.e. the mean absolute error between the output and corresponding pristine images, after each iteration (epoch) during training, along with the equivalent loss function obtained on a validation images set.

For the generation of the training set images we used the `pybinding` package, a Python package for tight-binding calculations, for the calculation of the energy- and spatially-dependent electronic structure of finite graphene nanosheets [48]. In this approach, the kernel polynomial method [49] is used to compute the projected density of states $c_{i,E}$ on each atomic site $i$ at energy $E$. Following Tersoff & Hamann [50], the STM tunneling current at a given sample bias $V_{Sample} = E/e$ ($e$ is the electron charge) is then computed for $V > 0$ as $I(x,y,z,V) \propto \sum_i \left(\sum_{E'=0}^{E} c_{i,E'}\right) \left|\pi_{z,i}(x,y,z)\right|^2$, where $\pi_{z,i}(x,y,z) \propto z e^{-r/\lambda}$ represents the value at $(x,y,z)$ of the $\pi_z$ orbital located at site $i$. Here, $r = \sqrt{(x-x_i)^2 + (y-y_i)^2 + (z-z_i)^2}$ defines the distance between the tip location and the carbon atom at site $i$. In the above, $\lambda \approx n a_0/Z$ = 0.018 nm for the given principal quantum number $n$ = 2, atomic number $Z$ = 6, and Bohr radius $a_0$. When the bias voltage $V_{Sample} < 0$, i.e. for tunneling from the sample to the tip, the sum over the projected density of states in STM current calculation above is replaced by $\sum_{E'=E}^{0} c_{i,E'}$. For computational efficiency, the STM training images were calculated for a constant height $z$ throughout each image. The lateral image size is chosen randomly between 1 and 4 nm with $64^2$ pixels.

In training and applying the CNN, we restrict ourselves to a particular material since it would be impractical to simulate STM images for all known materials. We chose monolayer and Bernal-stacked bilayer graphene as model materials. It should be noted that, because STM is surface sensitive, our model can in principle be applied to any graphitic material. While beyond the scope of our current work, the overall approach should be applicable to a larger range of materials, either for those with surfaces reasonably modeled by a similar tight-binding approach, or for materials requiring advanced electronic structure modeling using high-performance computation separately in the training step. Limitations are expected to arise, however, in more complex cases such as the presence of adsorbed molecules [51–54].

To make our denoising CNN as broadly applicable as possible, we train it on images that are as diverse as possible. To that end, we introduce the following variations, randomly chosen for each image: (*i*) Both monolayer and Bernal-stacked bilayer graphene are simulated. (*ii*) Dopants, simulated by changing the onsite energy to 10 eV at the position of the dopant, were randomly distributed. (*iii*) A hyperbolic background was added to the images. (*iv*) The lattice orientation, tip height $z$, and the sample bias between -0.5 to 0.5 V were randomly set. Moreover, two types of noise are included: (*v*) A Gaussian height noise with randomly chosen standard deviation $dz$ that is added onto each pixel, and (*vi*) a random horizontal spatial offset $dx$ is added for each line of the image, simulating a common line noise effect seen in STM images. The training images were first normalized to between -0.5 and 0.5, and the standard deviation dz of the added Gaussian height noise limited to a maximum value of 0.07 on this scale. In turn, the maximum value for the randomly chosen line noise $dx$ was 0.1 nm. A set of representative simulated STM images with ('corrupted') and without noise ('GT') used for training are shown in Supplemental Materials Fig. S1 [55]. Our code for producing the simulated images and for the ML model is provided on GitHub [56]. The production of simulated images and training of the CNN model were carried out on a modern computer with graphics processing unit in several hours, while denoising experimental images once the model is trained occurs on the order of seconds.

## III. RESULTS AND DISCUSSION

### A. Denoising of Simulated Images

Before its application to experimental data, we first analyze the performance of the ML-based model on simulated STM data and compare it to other denoising approaches. The top row of images in Fig. 2(a) shows a computed reference and corrupted image along with the results of denoising it with either the trained ML model or alternatively with Gaussian smoothing or singular value decomposition (SVD) filters. Gaussian smoothing consists of a convolution between the image and a 2D Gaussian profile with a fixed width $\sigma$. We chose $\sigma = 1.1$ pixels by maximizing the structural similarity index measure (SSIM) as a function of $\sigma$ (see Supplementary Materials). The SSIM is a widely used method for evaluating image similarity that compares the luminance, contrast, and image structure to a reference image [57]. In turn, SVD-based denoising uses the approach by Somnath et al. [58] It consists of tiling the $N \times N$ pixels image into $m \times m$ pixels windows, flattening the resulting 2D tile arrays onto 1D vectors and stacking them. A SVD decomposition is then applied to the resulting $(N - m + 1)^2 \times m^2$ matrix. The filtering consists of keeping only the first $t$ principal components. We set $m = 8$ and $t = 8$ for all SVD filtering presented in this work. Visual inspection in Fig. 2(a) of the results of these different denoising approaches reveals that the ML model performs well compared to the other methods. Both the height noise and line noise are almost entirely suppressed by the ML model, which is not the case for the other approaches. The contrast of the atomic lattice is also higher in the ML case.

Figure 2(c) plots the SSIM as a function of the image noise level, for each type of denoising approach. For producing the graph shown in Fig. 2(c) we generated ten sets of 200 corrupted images, with each set corresponding to a different value of the height noise $dz$ and line noise $dx \equiv dz$ plotted on the horizontal axis. Indeed, Fig. 2(c) quantitatively confirms the visual impression that the ML approach performs well both on an absolute scale and relative to the Gaussian smoothing and SVD methods at high noise levels. Importantly, even for low noise levels, ML denoising retains the SSIM while the other methods reduce the image quality for given fixed parameters. In the case of the ML approach, this consistent performance results from the analogy of minimizing the loss function (mean absolute error) during training and maximizing the SSIM.

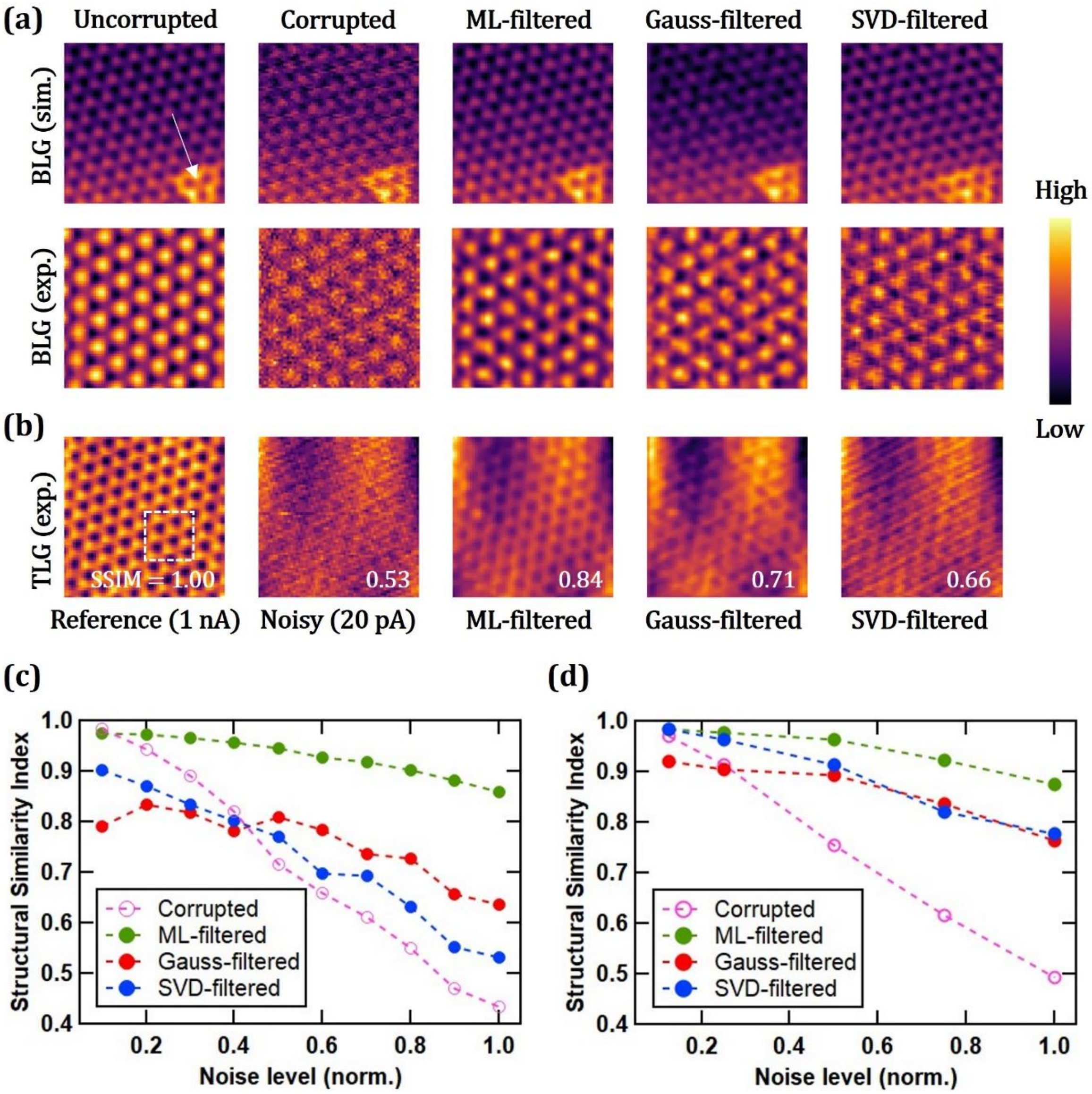

**Figure 2. Comparison of denoising STM images with different methods. (a)** STM images filtered with ML, Gaussian, and SVD-based denoising schemes, after corrupting a reference image with simulated height and line noise. The uncorrupted images represent bilayer graphene (BLG), based on a simulated STM image with N dopant (top row), and an experimentally-measured pristine BLG surface (second row). BLG model parameters are $V_{Sample} = -0.49$ eV, $z = 0.68$ nm, $dz = 0.041$, $dx = 0.107$ nm. **(b)** Comparison of filtering a noisy experimental STM image of Bernal-stacked trilayer graphene (TLG) via ML, Gaussian, and SVD methods. The noisy image was acquired with the same tip and area as the reference, yet at larger tip-sample distance i.e. low sample current. The structural similarity index measure (SSIM) with respect to the reference is indicated, as computed over a limited area (dashed box) to avoid artifacts the varying Moiré modulation contrast (see text; SSIM over the whole image are 1.00, 0.36, 0.55, 0.39, and 0.45 from left to right). **(c)** and **(d)** SSIM as a function of noise added onto (c) simulated STM images of mono- and bilayer graphene or onto (d) experimental BLG data as shown in the topmost panel.

## B. Experimental STM Image Denoising

We now show how our ML denoising method works on experimental data. The experimental STM images were obtained from scans carried out at liquid helium temperatures (4-5 K) without an applied external magnetic field. The images of pristine monolayer and bilayer graphene were obtained on graphene devices (see elsewhere for details on the making of these devices [59]). The images of doped and defected graphene were obtained on graphene grown on the carbon face of silicon carbide that was intentionally doped with nitrogen [60]. The experimental STM images were acquired at constant tunneling current, as customary to avoid crashing the tip, and thus represent the corresponding tip height variation. While the ML model is trained on constant-height data, the two modes generally produce similar images.

First, we study the performance based on experimental STM data where noise has been artificially added. Since the underlying high-quality STM image now serves as a reference image, a quantitative assessment of denoising can be obtained. The second row in Fig. 2(a) shows measured bilayer graphene (BLG) STM data (leftmost image) to which height and line noise is added before one of the three different denoising filters is applied. The dependence of the resulting SSIM on the added noise level is shown in Fig. 2(d). This confirms a strong overall SSIM enhancement from ≈0.49 to 0.87, fully comparable to the quantitative improvement obtained from denoising the simulated images discussed above. However, Gaussian and SVD filters also increase the SSIM significantly, which reduces the relative numerical advantage for ML-based denoising of these noise-corrupted experimental images.

In a second analysis, shown in Fig. 2(b), we recorded experimental STM images of Bernal-stacked trilayer graphene sample (TLG) at different noise levels. The noisy image was taken over the same area and with the same STM tip but at lower sample current (20 pA) compared to the high-quality reference (1 nA), resulting in a larger tip-sample distance. The denoised images using ML-, Gauss-, and SVD-filters are also shown: the apparent visual advantage of ML-denoising is confirmed by computing the SSIM between the denoised and reference images, resulting in an SSIM = 0.84 for the ML filter, compared to 0.71 and 0.66 for the Gauss and SVD filters, respectively. It should be noted that the TLG sample shows a long-range Moiré modulation due to the misalignment and lattice mismatch with the hBN substrate [61,62], whose relative contrast increases at larger tip-sample distance. To avoid artifacts on the SSIM values due to the changing Moiré contrast, we compute the SSIM over a limited area indicated in Fig. 2(b) (dashed box), with similar results obtained when the same-sized box is moved to other regions in the image. As the data shows, the advantage of ML filtering is thus not only visually apparent but also confirmed by the numerical similarity measure.

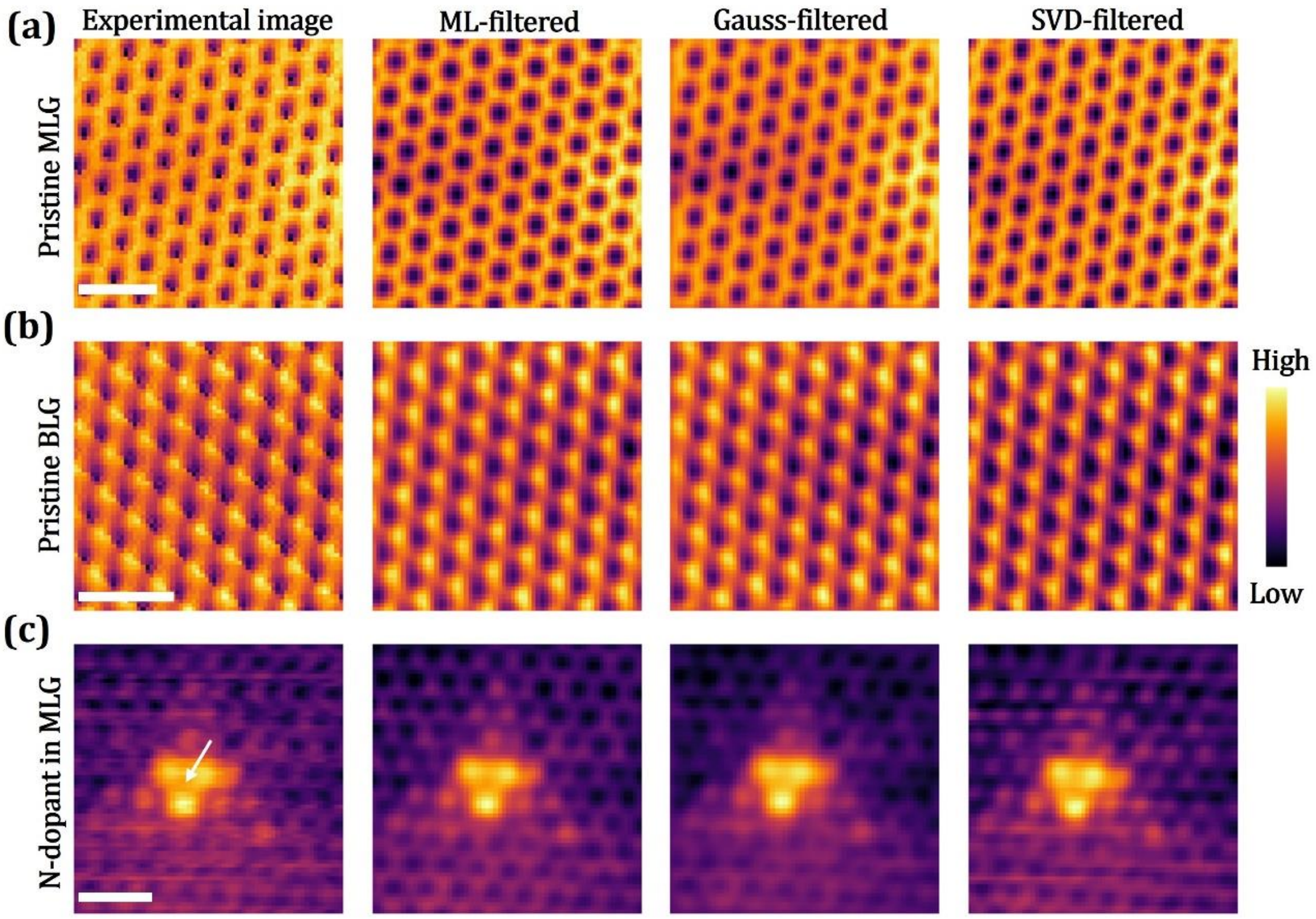

**Figure 3: Denoising of experimental STM image features simulated in the training**. Experimental STM images of **(a)** pristine monolayer graphene with $V_{sample} = -0.5$ V, **(b)** bilayer graphene with $V_{sample} = +0.3$ V and **(c)** monolayer graphene including an N dopant as indicated by the arrow ($V_{sample} = +0.5$ V). Scale bars are 0.5 nm. Results of applying the ML-based denoising to the experimental STM images are displayed in the second column. The results of Gaussian smoothing and SVD-based filtering are reported in the third and fourth columns, respectively.

Next, we discuss how our ML-based denoising model performs on images that contain (Fig. 3) or do not contain (Fig. 4) features simulated in the training set. The first column of Figs. 3(a) and 3(b) corresponds to experimentally measured STM images of pristine monolayer graphene (MLG) and BLG, respectively, and in Fig. 3(c) to monolayer graphene with a dopant (indicated by an arrow). Experiments and density functional theory calculations have shown that, for single nitrogen substitution in graphene, the highest electronic density is found on the three carbon atoms nearest neighbors to the nitrogen dopant [60,63], hence the three lobes clearly visible in Fig. 3(c). The second column in Fig. 3 shows the images denoised with our ML model. For comparison, the third and fourth columns show the same experimental images denoised with the Gaussian smoothing and the SVD filter, respectively. Note that the images in Figs. 3(a) and 3(b) also exhibit shape deformations due to a nonideal tip. Because uncorrupted reference images are inherently unavailable for these imperfect experimental images, a quantitative evaluation in terms of the SSIM is not possible. However, a visual inspection reveals that the ML approach performs well compared to the other filters. In particular, we note that the ML model more effectively removes the noise that appears as horizontal lines in the experimental images. The contrast corresponding to the atomic lattice also appears higher for the ML denoising [see Fig. 3(c)].

In Fig. 4, we show how our ML denoising performs on features that did not appear in the simulations used during training, where only single dopants were randomly incorporated. The first column displays three experimental images. The top image [Fig. 4(a)] is a 3N pyridine defect in graphene, which consists of a carbon vacancy accompanied by the replacement of its nearest neighbors with three nitrogen atoms in a pyridine configuration [64]. The center image [Fig. 4(b)] is the so-called “flower” defect in graphene, commonly observed in graphitic systems [65]. The bottom image [Fig. 4(c)] displays a grain boundary between two graphene domains, often found in graphene/SiC($000\bar{1}$) samples [66,67]. Visual inspection indicates a good performance of ML-based denoising (second column) when compared to the other two approaches (third and fourth columns). It should be noted that neither filter removes the short lateral lines (spike noise) in Fig. 4(c), which were not included in the training set for ML denoising. As in the cases discussed above, the contrast of the atomic lattice appears sharper for the ML-based method. That is especially true for the grain boundary defect [Fig. 4(c)], where the Gauss and SVD filters do not reproduce the graphene lattice that is visible on both sides of the grain boundary in the experimental and the ML-denoised images.

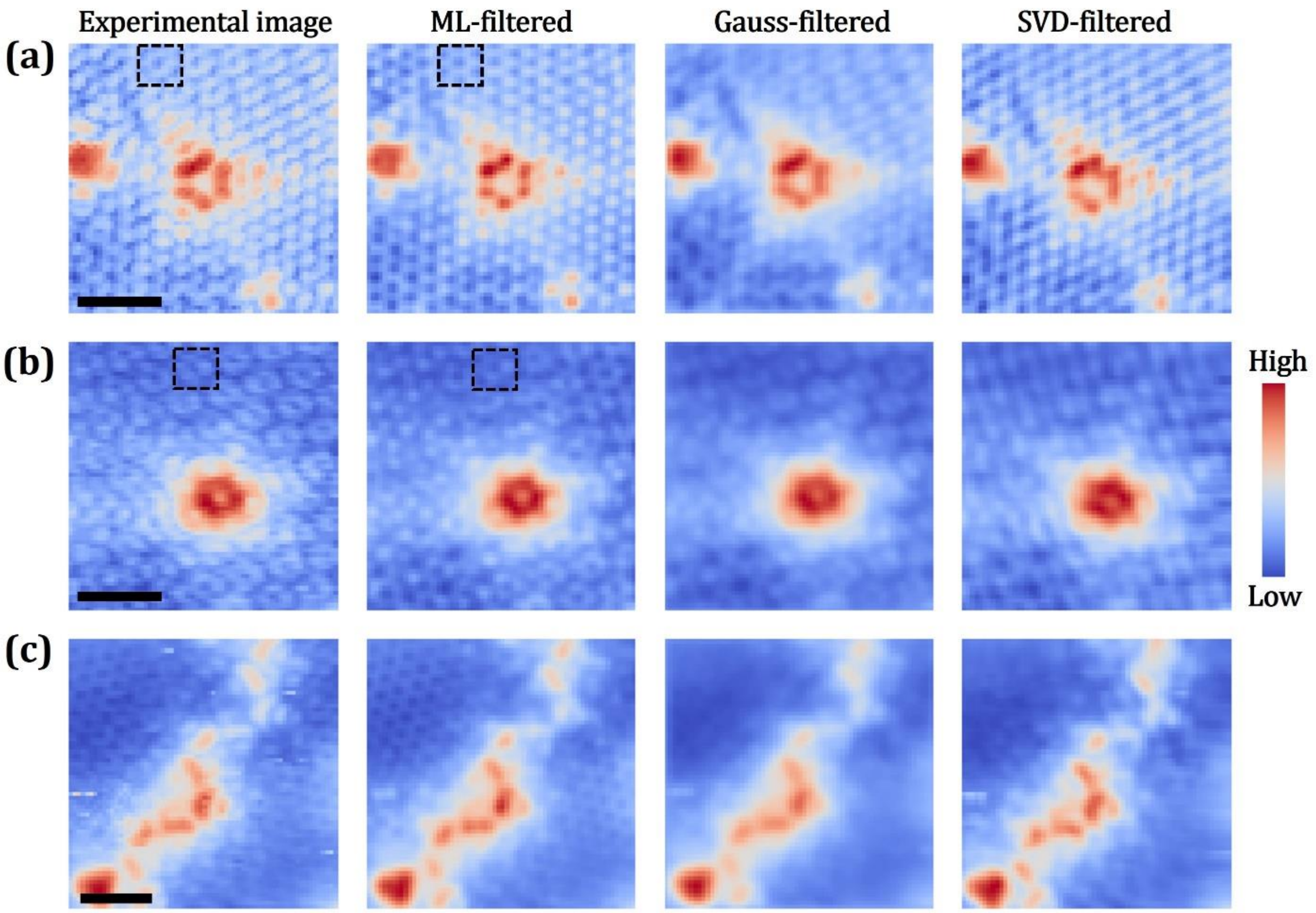

**Figure 4: Denoising of experimental monolayer graphene images with features not simulated in the training set**. **(a)** 3N pyridine defect ($V_{sample} = +0.5$ V). **(b)** 'Flower' defect ($V_{sample} = +0.5$ V). **(c)** Grain boundary between two graphene domains ($V_{sample} = +0.4$ V). Scale bars are 1 nm. Results of applying the ML-based denoising to the experimental images are displayed in the second column. The results of Gaussian smoothing and SVD filtering are reported in the third and fourth columns, respectively. The dashed black boxes in (a) and (b) encompass a donut forming the donut-like charge density wave feature (see text) which is partially smeared out by the ML-filtering.

A clear advantage of the ML approach over the other methods is the absence of parameters to be tuned once the model is trained. By comparison, other denoising methods including the Gaussian and SVD techniques entail parameters that need to be readjusted to different values depending on the type of image. In the images in Figs. 3(c) and 4(c), the same parameter $\sigma = 1.1$ pixel was used for Gaussian smoothing. This parameter works well for the dopant in Fig. 3(c), but it is not adapted for the grain boundary defect [Fig. 4(c)] and its associated high protrusion, as evidenced by the poorly resolved graphene lattice in the Gaussian-smoothed image. The same remark holds for the SVD approach, where two parameters must be set. The same parameters ($m = 8$ and $t = 8$, see above) were used in Figs. 3 and 4. The denoising effectiveness varies widely as evident from comparing the SVD-denoising of images in these figures.

### C. Extended STM Images

Because the training set of our ML model was limited to 64×64 pixels images and a lateral side length between 1–4 nm, it cannot be directly applied to larger images. While it is possible in principle to train the network on wider images with a higher pixel count, this rapidly increases the computational resources needed both for generation of the training set and for the training of the model. An alternative solution to apply ML denoising to larger images is to subdivide the large image into smaller images, apply the network to the smaller images, and then stitch the small images back together. This approach is illustrated in Fig. 5, on a $10 \times 10$ nm$^2$ image containing $256^2$ pixels. The image in Fig. 5(a) is subdivided into 16 images of $64^2$ pixels ($2.5 \times 2.5$ nm$^2$) onto which the ML model is applied and then assembled back together. The resulting denoised image is shown in Fig. 5(b). Except for faint lines at the borders between the sub-images, the resulting large ML-processed image (cf. insets in Fig. 5) is visually cleaner than the experimental image. While the z-scale is not directly conserved in the ML filter due to the inherent normalization, it can be well approximated in most cases by recording its range before normalization and reapplying it to the output image. This procedure has been applied in Fig. 5.

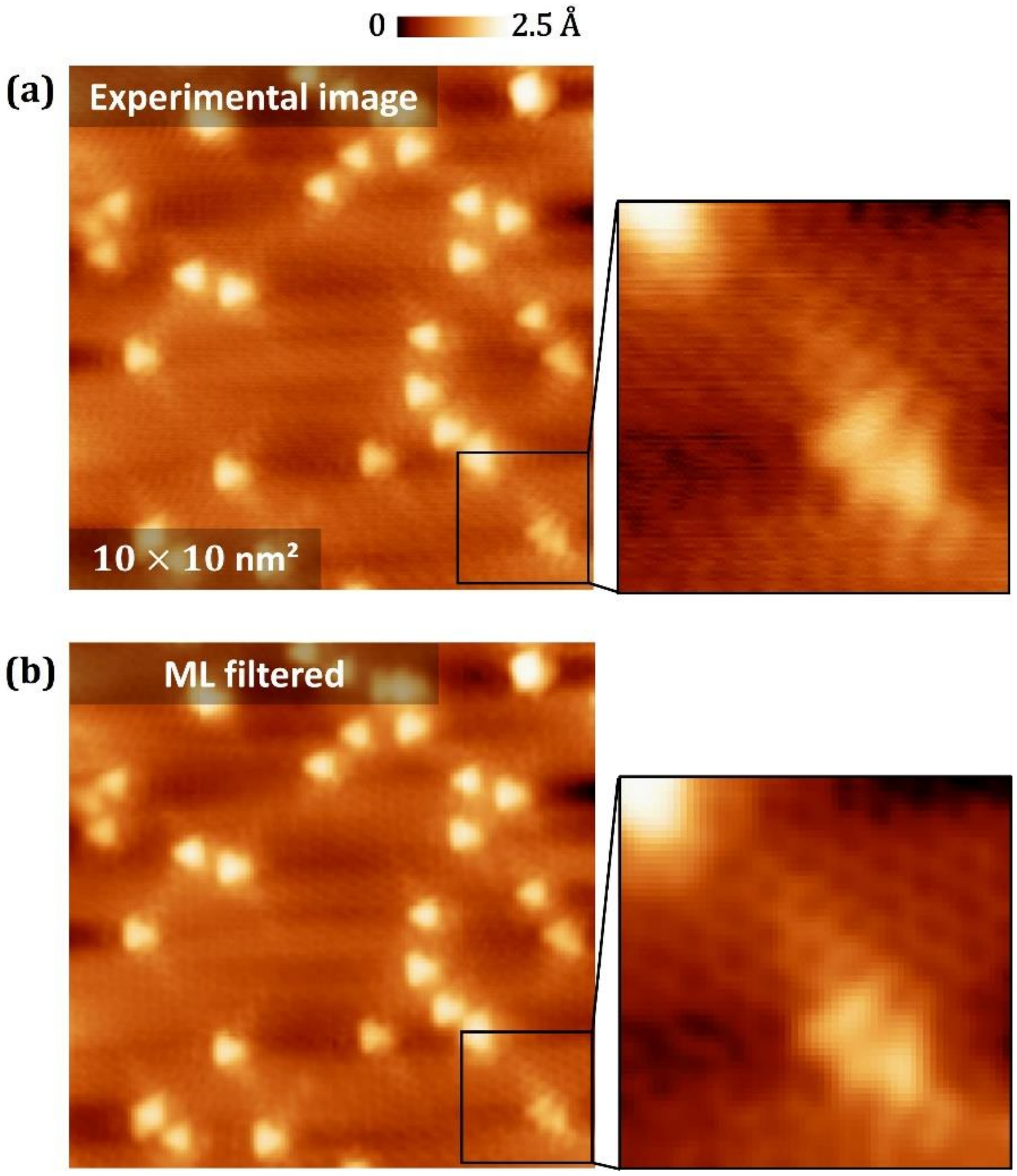

**Figure 5**: **ML-based denoising of extended images larger than the training images. (a)** $10 \times 10$ nm² image (256² pixels) showing monolayer graphene with N dopants. **(b)** ML-denoised image obtained by subdividing panel (a) into $4 \times 4$ images, applying the ML model to these images, and patching the ML-denoised images back together.

## D. Influence of the Training Set and Limitations

We now discuss how the inclusion of dopants, background, quasiparticle interference, and other features in the simulated STM training images affects the outcome of ML-based denoising. In Fig. 6, we show the same experimental images as in Fig. 4 and compare how the performance of the ML-denoising is affected by the inclusion or omission of features in the image set used for training the model. The case I (second column) was trained on simulated images containing only pristine monolayer and bilayer graphene. Perhaps unsurprisingly, this model poorly handles the filtering of images containing defects. The case II (third column) was trained on images of pristine monolayer and bilayer graphene, but with inclusion of quasiparticle interferences (QPI), but no dopant. Quasiparticle interferences in general are induced by charge-carrier scattering from defects such as dopants, step edges, or grain boundaries, and they produce standing-wave patterns [68,69]. For the simulated images in the training set, QPI is inherently included in the tight-binding calculation due to scattering off the boundaries of the simulated graphene nanosheet. In this case, it becomes prominent only when nanosheet size is small (cf. Supplementary Fig. S3 and refs. [70,71]). Such simulated images with evident QPI can be seen e.g. in the zig-zag type patterns in Figs. S1(a) and S1(j). Surprisingly, training with such QPI-pattern images alone is sufficient for the model to perform relatively well on denoising of images that include defects, as evident from Fig. 6. Only the case of the grain boundary [Fig. 6(c)] is poorly handled by this approach. The case III in Fig. 6 combines images containing dopants in the training set and results in the model being able to handle more complex defects such as the grain boundary.

Finally, we discuss a feature that is partially smeared out by our ML-filtering. As evidenced in Fig. 4, a donut-like pattern is present, highlighted by the dotted box in Figs. 4(a) and 4(b). This donut-like pattern arises from a charge-density wave due to a Kékulé distortion whose periodicity is $\sqrt{3}$ larger than that of the graphene lattice. Similar STM signatures have been observed recently as a broken symmetry state in charge neutral graphene, in intense magnetic fields [72,73], and can be reproduced with a tight-binding model if one assumes that valleys and sublattices are locked i.e. that charge carriers in one of the two inequivalent valleys of graphene comprise only carriers on one of the two sublattices [72]. Similar Kékulé distortions have been observed by STM, in the absence of magnetic field, in Li-intercalated graphene [74] and in graphene grown epitaxially on copper [75]. However, these previously reported STM signatures do not match the distinct donut-like pattern that we report here for defected graphene, which corresponds to a certain mixture of K and K’ states [72].

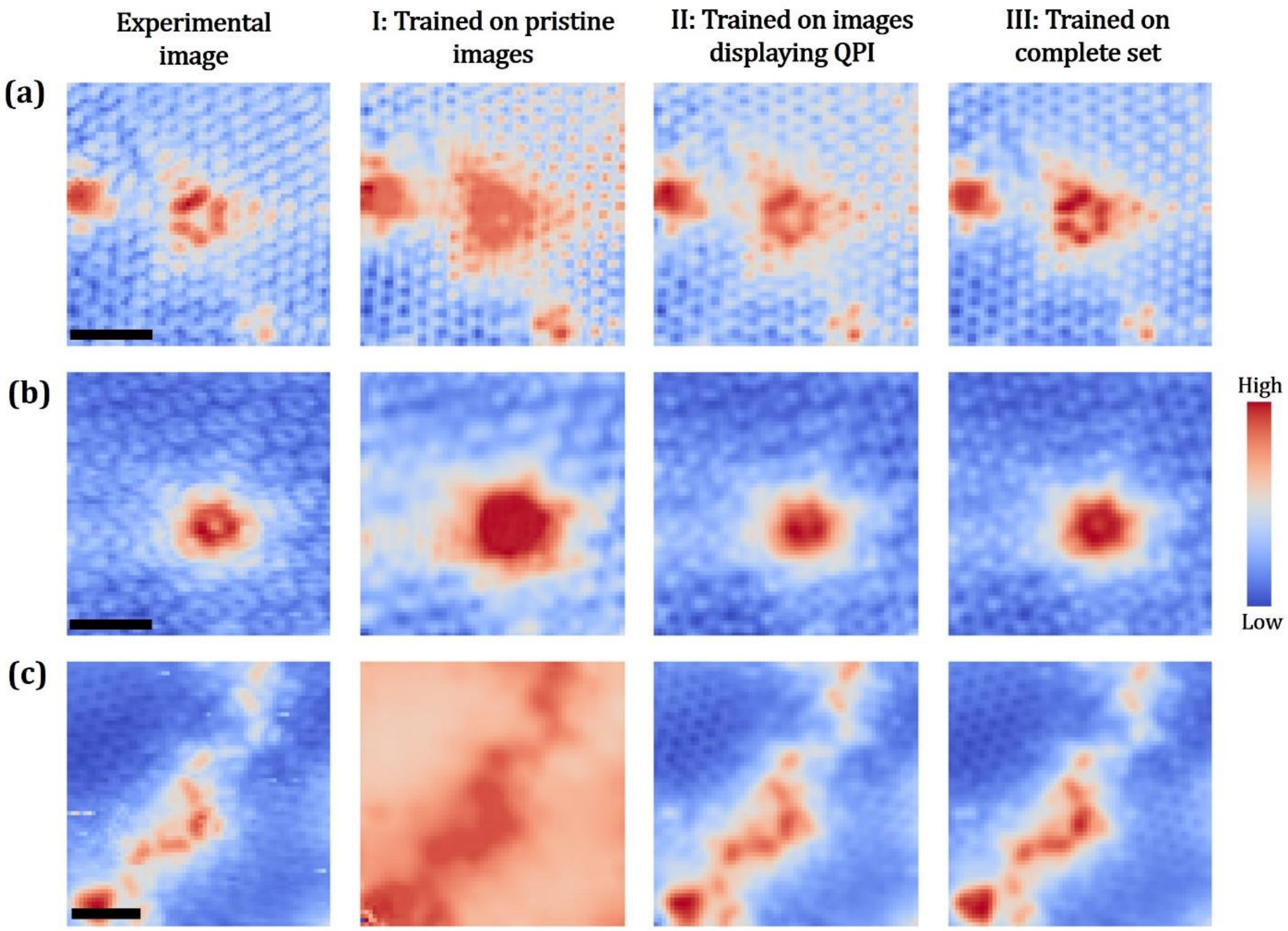

**Figure 6**: **Influence of the introduction of various features in the simulated images used for training the ML model. (a)-(c)** Experimental images as in Fig. 4. Scale bars are 1 nm. Case I: results from the ML-filter model trained only with simulated pristine monolayer and bilayer graphene STM images. Case II: results from the ML model trained with an image set that includes simulated images displaying quasiparticle interference patterns (see text and Suppl. Materials) but no dopant. Case III: results from the ML model trained with images containing all the features (same as Fig. 4, second column; see section II for details on all the included features).

The tight-binding model underlying the simulated STM images used to train our model does not reproduce the donut-like charge density wave state, since we had no a priori reason to assume any valley-sublattice locking. This lack of inclusion in the training set explains why the donut-like pattern tends to be partially smeared out by the ML-based denoising model as evident from the second column images in Figs. 4(a) and 4(b). This issue illustrates well an intrinsic limitation from the bias induced by a non-representative training set in machine learning [9,76–78]. While the feature is not completely smeared out [cf. Fig. 4(b)] this demonstrates that care should be taken when applying ML-based models to experimental data in fundamental physics experiments if previously unexplored questions are tackled, emphasizing the need for raw images in such cases.

Overall, ML-based denoising can provide multiple benefits to STM applications. As discussed earlier, the significant noise reduction achieved can already increase the level of confidence when interpreting data visually. Importantly, performant denoising can also boost the efficiency of subsequent ML architectures such as defect identification or automated scanning, which can be of use e.g. in the efficient development and analysis of novel 2D electronic materials or defect-based quantum information platforms. Perhaps the most important benefit of efficient denoising lies in increasing the throughput of STM experiments, rendering usable images that would normally be discarded. This allows for greater scanning speed and a choice of parameters, such as low tunneling current, that are less likely to produce tip degradation.

## IV. CONCLUSIONS

In summary, we have introduced a machine learning model for denoising STM images of graphitic systems which is trained on simulated images. The wide variety of simulated images in the training set allows us to apply the model to a broad range of experimental images. We also demonstrate that by tiling the model can be applied to denoise images that are larger than those used in training while retaining a rapid computational speed. The relevance of including different features in the simulated training set was analyzed, revealing that the inclusion of QPI patterns alone results in surprisingly good ML denoising. Finally, limitations of ML-based denoising were illustrated in the case of an unexpected charge-density wave distortion. The ML denoising approach presented here is expected to be an effective tool for facilitating the visual analysis of STM images and for enhancing subsequent machine learning techniques applied to defect detection and classification, atom manipulation, and scan automation.

**ACKNOWLEDGMENTS**. J.V.J. acknowledges support from the National Science Foundation under Grant No. DMR-1753367 and the Army Research Office under Contract No. W911NF-17-1-0473. K.W. and T.T. acknowledge support from the Elemental Strategy Initiative conducted by the MEXT, Japan, Grant No. JPMXP0112101001 and JSPS KAKENHI Grant No. JP20H00354.

## REFERENCES

[1] G. Binnig, H. Rohrer, Ch. Gerber, and E. Weibel, *Surface Studies by Scanning Tunneling Microscopy*, Phys Rev Lett **49**, 57 (1982).

[2] C. Julian Chen, *Introduction to Scanning Tunneling Microscopy*, Third (Oxford University Press, USA, 2021).

[3] M. F. Crommie, C. P. Lutz, and D. M. Eigler, *Confinement of Electrons to Quantum Corrals on a Metal Surface*, Science (1979) **262**, 218 (1993).

[4] M. F. Crommie, C. P. Lutz, and D. M. Eigler, *Imaging Standing Waves in a Two-Dimensional Electron Gas*, Nature **363**, 524 (1993).

[5] Ø. Fischer, M. Kugler, I. Maggio-Aprile, C. Berthod, and C. Renner, *Scanning Tunneling Spectroscopy of High-Temperature Superconductors*, Rev Mod Phys **79**, 353 (2007).

[6] E. Bedolla, L. C. Padierna, and R. Castañeda-Priego, *Machine Learning for Condensed Matter Physics*, Journal of Physics: Condensed Matter **33**, 053001 (2020).

[7] R. Pederson, B. Kalita, and K. Burke, *Machine Learning and Density Functional Theory*, Nature Reviews Physics **4**, 357 (2022).

[8] G. Carleo, I. Cirac, K. Cranmer, L. Daudet, M. Schuld, N. Tishby, L. Vogt-Maranto, and L. Zdeborová, *Machine Learning and the Physical Sciences*, Rev Mod Phys **91**, 45002 (2019).

[9] J. Yano, K. J. Gaffney, J. Gregoire, L. Hung, A. Ourmazd, J. Schrier, J. A. Sethian, and F. M. Toma, *The Case for Data Science in Experimental Chemistry: Examples and Recommendations*, Nat Rev Chem **6**, 357 (2022).

[10] M. Ziatdinov, C. Y. Wong, and S. v. Kalinin, *Finding Simplicity: Unsupervised Discovery of Features, Patterns, and Order Parameters via Shift-Invariant Variational Autoencoders*, (2021).

[11] Y. Zhang, A. Mesaros, K. Fujita, S. D. Edkins, M. H. Hamidian, K. Ch'ng, H. Eisaki, S. Uchida, J. C. Séamus Davis, E. Khatami and E.-A. Kim, *Machine Learning in Electronic-Quantum-Matter Imaging Experiments*, Nature **570**, 484 (2019).

[12] K. M. Roccapriore, Q. Zou, L. Zhang, R. Xue, J. Yan, M. Ziatdinov, M. Fu, D. G. Mandrus, M. Yoon, B. G. Sumpter, Z. Gai, and S. V. Kalinin, *Revealing the Chemical Bonding in Adatom Arrays via Machine Learning of Hyperspectral Scanning Tunneling Spectroscopy Data*, ACS Nano **15**, 11806 (2021).

[13] M. Rashidi, J. Croshaw, K. Mastel, M. Tamura, H. Hosseinzadeh, and R. A. Wolkow, *Deep Learning-Guided Surface Characterization for Autonomous Hydrogen Lithography*, Mach Learn Sci Technol **1**, 025001 (2020).

[14] M. Ziatdinov, U. Fuchs, J. H. G. Owen, J. N. Randall, and S. v. Kalinin, *Robust Multi-Scale Multi-Feature Deep Learning for Atomic and Defect Identification in Scanning Tunneling Microscopy on H-Si(100) 2x1 Surface*, (2020).

[15] M. Rashidi and R. A. Wolkow, *Autonomous Scanning Probe Microscopy in Situ Tip Conditioning through Machine Learning*, ACS Nano **12**, 5185 (2018).

[16] A. Krull, P. Hirsch, C. Rother, A. Schiffrin, and C. Krull, *Artificial-Intelligence-Driven Scanning Probe Microscopy*, Commun Phys **3**, 54 (2020).

[17] K. Zhang, W. Zuo, and L. Zhang, *FFDNet: Toward a Fast and Flexible Solution for CNN-Based Image Denoising*, IEEE Transactions on Image Processing **27**, 4608 (2018).

[18] P. Liu, H. Zhang, K. Zhang, L. Lin, and W. Zuo, *Multi-Level Wavelet-CNN for Image Restoration*, in *Proceedings of the IEEE Conference on Computer Vision and Pattern Recognition Workshops* (2018), pp. 773–782.

[19] X. Jia, S. Liu, X. Feng, and L. Zhang, *Focnet: A Fractional Optimal Control Network for Image Denoising*, in *Proceedings of the IEEE/CVF Conference on Computer Vision and Pattern Recognition* (2019), pp. 6054–6063.

[20] S. Lefkimmiatis, *Universal Denoising Networks: A Novel CNN Architecture for Image Denoising*, in *Proceedings of the IEEE Conference on Computer Vision and Pattern Recognition* (2018), pp. 3204–3213.

[21] I. Kligvasser, T. R. Shaham, and T. Michaeli, *Xunit: Learning a Spatial Activation Function for Efficient Image Restoration*, in *Proceedings of the IEEE Conference on Computer Vision and Pattern Recognition* (2018), pp. 2433–2442.

[22] S. Lefkimmiatis, *Non-Local Color Image Denoising with Convolutional Neural Networks*, in *Proceedings of the IEEE Conference on Computer Vision and Pattern Recognition* (2017), pp. 3587–3596.

[23] V. Jain and S. Seung, *Natural Image Denoising with Convolutional Networks*, Adv Neural Inf Process Syst **21**, (2008).

[24] Y. Tai, J. Yang, X. Liu, and C. Xu, *Memnet: A Persistent Memory Network for Image Restoration*, in *Proceedings of the IEEE International Conference on Computer Vision* (2017), pp. 4539–4547.

[25] K. Zhang, W. Zuo, Y. Chen, D. Meng, and L. Zhang, *Beyond a Gaussian Denoiser: Residual Learning of Deep Cnn for Image Denoising*, IEEE Transactions on Image Processing **26**, 3142 (2017).

[26] X. Mao, C. Shen, and Y.-B. Yang, *Image Restoration Using Very Deep Convolutional Encoder-Decoder Networks with Symmetric Skip Connections*, Adv Neural Inf Process Syst **29**, 2802 (2016).

[27] Y. Chen and T. Pock, *Trainable Nonlinear Reaction Diffusion: A Flexible Framework for Fast and Effective Image Restoration*, IEEE Trans Pattern Anal Mach Intell **39**, 1256 (2016).

[28] B. Manifold, E. Thomas, A. T. Francis, A. H. Hill, and D. Fu, *Denoising of Stimulated Raman Scattering Microscopy Images via Deep Learning*, Biomed Opt Express **10**, 3860 (2019).

[29] Y. Kim, D. Oh, S. Huh, D. Song, S. Jeong, J. Kwon, M. Kim, D. Kim, H. Ryu, J. Jung et al., *Deep Learning-Based Statistical Noise Reduction for Multidimensional Spectral Data*, Review of Scientific Instruments **92**, 073901 (2021).

[30] J. Tersoff and D. R. Hamann, *Theory and Application for the Scanning Tunneling Microscope*, Phys Rev Lett **50**, 1998 (1983).

[31] Y. Zhang, V. W. Brar, F. Wang, C. Girit, Y. Yayon, M. Panlasigui, A. Zettl, and M. F. Crommie, *Giant Phonon-Induced Conductance in Scanning Tunnelling Spectroscopy of Gate-Tunable Graphene*, Nat Phys **4**, 627 (2008).

[32] S. Ernst, S. Wirth, M. Rams, V. Dolocan, and F. Steglich, *Tip Preparation for Usage in an Ultra-Low Temperature UHV Scanning Tunneling Microscope*, Sci Technol Adv Mater **8**, 347 (2007).

[33] J. G. Rodrigo and S. Vieira, *STM Study of Multiband Superconductivity in NbSe2 Using a Superconducting Tip*, Physica C Supercond **404**, 306 (2004).

[34] I. Ekvall, E. Wahlström, D. Claesson, H. Olin, and E. Olsson, *Preparation and Characterization of Electrochemically Etched W Tips for STM*, Meas Sci Technol **10**, 11 (1999).

[35] M. Stocker, H. Pfeifer, and B. Koslowski, *Calibration of Tip and Sample Temperature of a Scanning Tunneling Microscope Using a Superconductive Sample*, Journal of Vacuum Science & Technology A **32**, 031605 (2014).

[36] S. Mohan, R. Manzorro, J. L. Vincent, B. Tang, D. Y. Sheth, E. P. Simoncelli, D. S. Matteson, P. A. Crozier, and C. Fernandez-Granda, *Deep Denoising for Scientific Discovery: A Case Study in Electron Microscopy*, ArXiv Preprint ArXiv:2010.12970 (2020).

[37] Y. Zhang, Y. Zhu, E. Nichols, Q. Wang, S. Zhang, C. Smith, and S. Howard, *A Poisson-Gaussian Denoising Dataset With Real Fluorescence Microscopy Images*, in *Proceedings of the IEEE/CVF Conference on Computer Vision and Pattern Recognition (CVPR)* (2019).

[38] M. Prakash, M. Lalit, P. Tomancak, A. Krul, and F. Jug, *Fully Unsupervised Probabilistic Noise2Void*, in *2020 IEEE 17th International Symposium on Biomedical Imaging (ISBI)* (2020), pp. 154–158.

[39] W. Khademi, S. Rao, C. Minnerath, G. Hagen, and J. Ventura, *Self-Supervised Poisson-Gaussian Denoising*, in *2021 IEEE Winter Conference on Applications of Computer Vision (WACV)* (2021), pp. 2130–2138.

[40] S. Laine, T. Karras, J. Lehtinen, and T. Aila, *High-Quality Self-Supervised Deep Image Denoising*, in *Advances in Neural Information Processing Systems*, edited by H. Wallach, H. Larochelle, A. Beygelzimer, F. d Alché-Buc, E. Fox, and R. Garnett, Vol. 32 (Curran Associates, Inc., 2019).

[41] A. Krull, T.-O. Buchholz, and F. Jug, *Noise2Void - Learning Denoising from Single Noisy Images*, (2018).

[42] J. Batson and L. Royer, *Noise2Self: Blind Denoising by Self-Supervision*, (2019).

[43] J.-F. Ge, M. Ovadia, and J. E. Hoffman, *Achieving Low Noise in Scanning Tunneling Spectroscopy*, Review of Scientific Instruments **90**, 101401 (2019).

[44] F. Chollet and Others, *Keras*, https://github.com/fchollet/keras.

[45] M. Abadi, A. Agarwal, P. Barham, E. Brevdo, Z. Chen, C. Citro, G. S. Corrado, A. Davis, J. Dean, M. Devin et al., *TensorFlow: Large-Scale Machine Learning on Heterogeneous Systems*.

[46] O. Ronneberger, P. Fischer, and T. Brox, *U-Net: Convolutional Networks for Biomedical Image Segmentation*, in *Medical Image Computing and Computer-Assisted Intervention – MICCAI 2015*, edited by N. Navab, J. Hornegger, W. M. Wells, and A. F. Frangi (Springer International Publishing, Cham, 2015), pp. 234–241.

[47] D. P. Kingma and J. Ba, *Adam: A Method for Stochastic Optimization*.

[48] D. Moldovan, M. Anđelković, and F. Peeters, *Pybinding v0.9.5: A Python Package for Tight-Binding Calculations*.

[49] A. Weiße, G. Wellein, A. Alvermann, and H. Fehske, *The Kernel Polynomial Method*, Rev Mod Phys **78**, 275 (2006).

[50] J. Tersoff and D. R. Hamann, *Theory of the Scanning Tunneling Microscope*, Phys Rev B **31**, 805 (1985).

[51] K. Bairagi et al., *Molecular-Scale Dynamics of Light-Induced Spin Cross-over in a Two-Dimensional Layer*, Nat Commun **7**, 12212 (2016).

[52] C. Fourmental et al., *Importance of Epitaxial Strain at a Spin-Crossover Molecule–Metal Interface*, J Phys Chem Lett **10**, 4103 (2019).

[53] V. D. Pham, S. Ghosh, F. Joucken, M. Pelaez-Fernandez, V. Repain, C. Chacon, A. Bellec, Y. Girard, R. Sporken, S. Rousset et al., *Selective Control of Molecule Charge State on Graphene Using Tip-Induced Electric Field and Nitrogen Doping*, NPJ 2D Mater Appl **3**, 5 (2019).

[54] D. Li, C. Barreteau, S. L. Kawahara, J. Lagoute, C. Chacon, Y. Girard, S. Rousset, V. Repain, and A. Smogunov, *Symmetry-Selected Spin-Split Hybrid States in ${\mathrm{C}}_{60}/\text{ferromagnetic}$ Interfaces*, Phys Rev B **93**, 85425 (2016).

[55] *See Supplemental Material at [...] for Representative Simulated STM Images, Choice of Gaussian Filter Width, and Quasiparticle Interference Generation in Simulated STM Images*.

[56] *Https://Github.Com/Fjoucken/Denoise_STM*, (unpublished).

[57] Z. Wang, A. C. Bovik, H. R. Sheikh, and E. P. Simoncelli, *Image Quality Assessment: From Error Visibility to Structural Similarity*, IEEE Transactions on Image Processing **13**, 600 (2004).

[58] S. Somnath, C. R. Smith, S. v Kalinin, M. Chi, A. Borisevich, N. Cross, G. Duscher, and S. Jesse, *Feature Extraction via Similarity Search: Application to Atom Finding and Denoising in Electron and Scanning Probe Microscopy Imaging*, Adv Struct Chem Imaging **4**, 3 (2018).

[59] F. Joucken, C. Bena, Z. Ge, E. A. Quezada-Lopez, S. Pinon, V. Kaladzhyan, T. Tanigushi, K. Watanabe, and J. Velasco, *Direct Visualization of Native Defects in Graphite and Their Effect on the Electronic Properties of Bernal-Stacked Bilayer Graphene*, Nano Lett **21**, 7100 (2021).

[60] F. Joucken, Y. Tison, J. Lagoute, J. Dumont, D. Cabosart, B. Zheng, V. Repain, C. Chacon, Y. Girard, A. Rafael Botello-Mendez et al., *Localized State and Charge Transfer in Nitrogen-Doped Graphene*, Phys Rev B Condens Matter Mater Phys **85**, 161408 (2012).

[61] F. Joucken, E. A. Quezada-López, J. Avila, C. Chen, J. L. Davenport, H. Chen, K. Watanabe, T. Taniguchi, M. C. Asensio, and J. Velasco, *Nanospot Angle-Resolved Photoemission Study of Bernal-Stacked Bilayer Graphene on Hexagonal Boron Nitride: Band Structure and Local Variation of Lattice Alignment*, Phys Rev B **99**, 161406 (2019).

[62] J. Xue, J. Sanchez-Yamagishi, D. Bulmash, P. Jacquod, A. Deshpande, K. Watanabe, T. Taniguchi, P. Jarillo-Herrero, and B. J. LeRoy, *Scanning Tunnelling Microscopy and Spectroscopy of Ultra-Flat Graphene on Hexagonal Boron Nitride*, Nat Mater **10**, 282 (2011).

[63] L. Zhao, R. He, K. T. Rim, T. Schiros, K. S. Kim, H. Zhou, C. Gutiérrez, S. P. Chockalingam, C. J. Arguello, L. Pálová et al., *Visualizing Individual Nitrogen Dopants in Monolayer Graphene*, Science (1979) **333**, 999 (2011).

[64] Y. Tison, J. Lagoute, V. Repain, C. Chacon, Y. Girard, S. Rousset, F. Joucken, D. Sharma, L. Henrard, H. Amara et al., *Electronic Interaction between Nitrogen Atoms in Doped Graphene*, ACS Nano **9**, 670 (2015).

[65] E. Cockayne, G. M. Rutter, N. P. Guisinger, J. N. Crain, P. N. First, and J. A. Stroscio, *Grain Boundary Loops in Graphene*, Phys Rev B **83**, 195425 (2011).

[66] F. Varchon, P. Mallet, L. Magaud, and J.-Y. Veuillen, *Rotational Disorder in Few-Layer Graphene Films on 6H-SiC(000-1): A Scanning Tunneling Microscopy Study*, Phys Rev B **77**, 165415 (2008).

[67] Y. Tison, J. Lagoute, V. Repain, C. Chacon, Y. Girard, F. Joucken, R. Sporken, F. Gargiulo, O. v Yazyev, and S. Rousset, *Grain Boundaries in Graphene on SiC(000$\bar{1}$) Substrate*, Nano Lett **14**, 6382 (2014).

[68] M. F. Crommie, C. P. Lutz, and D. M. Eigler, *Imaging Standing Waves in a Two-Dimensional Electron Gas*, Nature **363**, 524 (1993).

[69] J. E. Hoffman, *Spectroscopic Scanning Tunneling Microscopy Insights into Fe-Based Superconductors*, Reports on Progress in Physics **74**, 124513 (2011).

[70] A. Mahmood, P. Mallet, and J.-Y. Veuillen, *Quasiparticle Scattering off Phase Boundaries in Epitaxial Graphene*, Nanotechnology **23**, 055706 (2012).

[71] J. Tesch, P. Leicht, F. Blumenschein, L. Gragnaniello, A. Bergvall, T. Löfwander, and M. Fonin, *Impurity Scattering and Size Quantization Effects in a Single Graphene Nanoflake*, Phys Rev B **95**, 75429 (2017).

[72] L. Xiaomeng, F. Gelareh, C. Cheng-Li, P. Zlatko, W. Kenji, T. Takashi, Z. M. P, and Y. Ali, *Visualizing Broken Symmetry and Topological Defects in a Quantum Hall Ferromagnet*, Science (1979) **375**, 321 (2022).

[73] A. Coissard, D. Wander, H. Vignaud, A. G. Grushin, C. Repellin, K. Watanabe, T. Taniguchi, F. Gay, C. B. Winkelmann, H. Courtois, and H. Sellier, *Imaging Tunable Quantum Hall Broken-Symmetry Orders in Graphene*, Nature **605**, 51 (2022).

[74] C. Bao, H. Zhang, T. Zhang, X. Wu, L. Luo, S. Zhou, Q. Li, Y. Hou, W. Yao, L. Liu et al., *Experimental Evidence of Chiral Symmetry Breaking in Kekulé-Ordered Graphene*, Phys Rev Lett **126**, 206804 (2021).

[75] C. Gutiérrez, C.-J. Kim, L. Brown, T. Schiros, D. Nordlund, E. B. Lochocki, K. M. Shen, J. Park, and A. N. Pasupathy, *Imaging Chiral Symmetry Breaking from Kekulé Bond Order in Graphene*, Nat Phys **12**, 950 (2016).

[76] N. Mehrabi, F. Morstatter, N. Saxena, K. Lerman, and A. Galstyan, *A Survey on Bias and Fairness in Machine Learning*, ACM Comput. Surv. **54**, (2021).

[77] G. E. Karniadakis, I. G. Kevrekidis, L. Lu, P. Perdikaris, S. Wang, and L. Yang, *Physics-Informed Machine Learning*, Nature Reviews Physics **3**, 422 (2021).

[78] A. Torralba and A. A. Efros, *Unbiased Look at Dataset Bias*, in *CVPR 2011* (2011), pp. 1521–1528.

# Supplemental Materials for:

# Denoising Scanning Tunneling Microscopy Images of Graphene with Supervised Machine Learning

Frédéric Joucken[1], John L. Davenport[2], Zhehao Ge[2], Eberth A. Quezada-Lopez[2], Takashi Taniguchi[3], Kenji Watanabe[4], Jairo Velasco Jr.[2], Jérôme Lagoute[5], and Robert A. Kaindl[1]

[1]*Department of Physics, Arizona State University, Tempe, AZ 85287, USA*

[2]*Department of Physics, University of California, Santa Cruz, CA 95064, USA*

[3]*International Center for Materials Nanoarchitectonics, National Institute for Materials Science, 1-1 Namiki, Tsukuba 305-0044, Japan*

[4]*Research Center for Functional Materials, National Institute for Materials Science, 1-1 Namiki, Tsukuba 305-0044, Japan*

[5]*Laboratoire Matériaux et Phénomènes Quantiques, UMR 7162, Université Paris Diderot – Paris 7, Sorbonne Paris Cité, CNRS, UMR 7162 case courrier 7021, 75205 Paris 13, France*

Figure S1 shows representative simulated STM images used for training. The corresponding most important parameters are indicated below each image: whether it is monolayer graphene or bilayer graphene (MLG or BLG), the lateral size of the image, the energy $E$, the tip height $z$ (arb. u.), and the two noise parameters $dz$ and $dx$. $dz$ is the standard deviation of the Gaussian noise added on the image (in arb. u.) and $dx$ is the upper limit for the randomly chosen lateral offset between each horizontal lines of the image (in nm). Randomly distributed dopants are also visible on some images.

In Fig. S2, the evolution of the structural similarity index measure (SSIM) is shown as a function of the standard deviation of the Gaussian filter (in pixels) applied to corrupted simulated images. The SSIM has been computed for 200 images and the error bars correspond to the standard deviation of the ensemble of the 200 images around the average. The maximum SSIM is observed around 1.1, hence our choice of $\sigma = 1.1$ for the Gaussian filter used for the data presented in the main text.

Figure S3 illustrates the method to induce quasiparticle interference (QPI) patterns in the simulated STM image. To introduce QPI patterns, it suffices to consider graphene nanosheets which are only slightly larger than the area considered to produce the STM image (located at the center of the graphene nanosheet). The QPI are then produced by the scattering of the quasiparticle off the edges of the simulated nanosheet. [1,2] In Fig. S3(a), the nanosheet is 2 nm larger than the STM image (4 nm vs. 2 nm). In Fig. S3(b), the nanosheet is 10 nm larger than the STM image (12 nm vs. 2 nm). The other parameters are the same. One can see that QPI patterns are prominent in Fig. S3(a), whereas almost no QPI pattern is visible in Fig. S3(b).

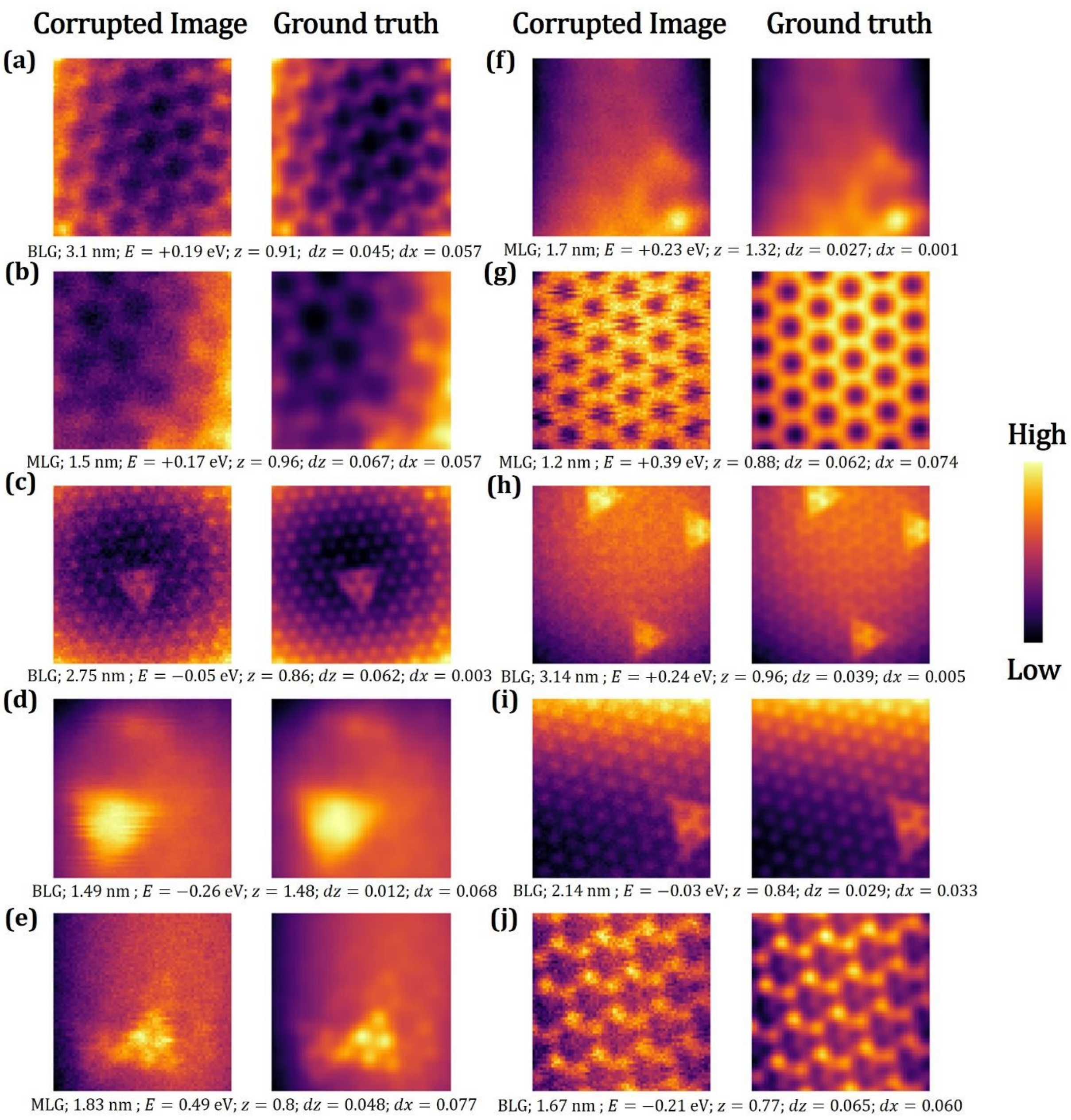

**Figure S1. Representative sample of simulated STM images used for training the model**. Below each image, the main parameters are indicated: whether it is monolayer graphene or bilayer graphene (MLG or BLG), the lateral size of the image, the energy E, the tip height z, and the two noise parameters dz and dx. dz is the standard deviation of the Gaussian noise added on the image (in arb. u.) and dx is the upper limit for the randomly chosen lateral offset between each horizontal lines of the image (in nm). Randomly distributed dopants are also visible on some images.

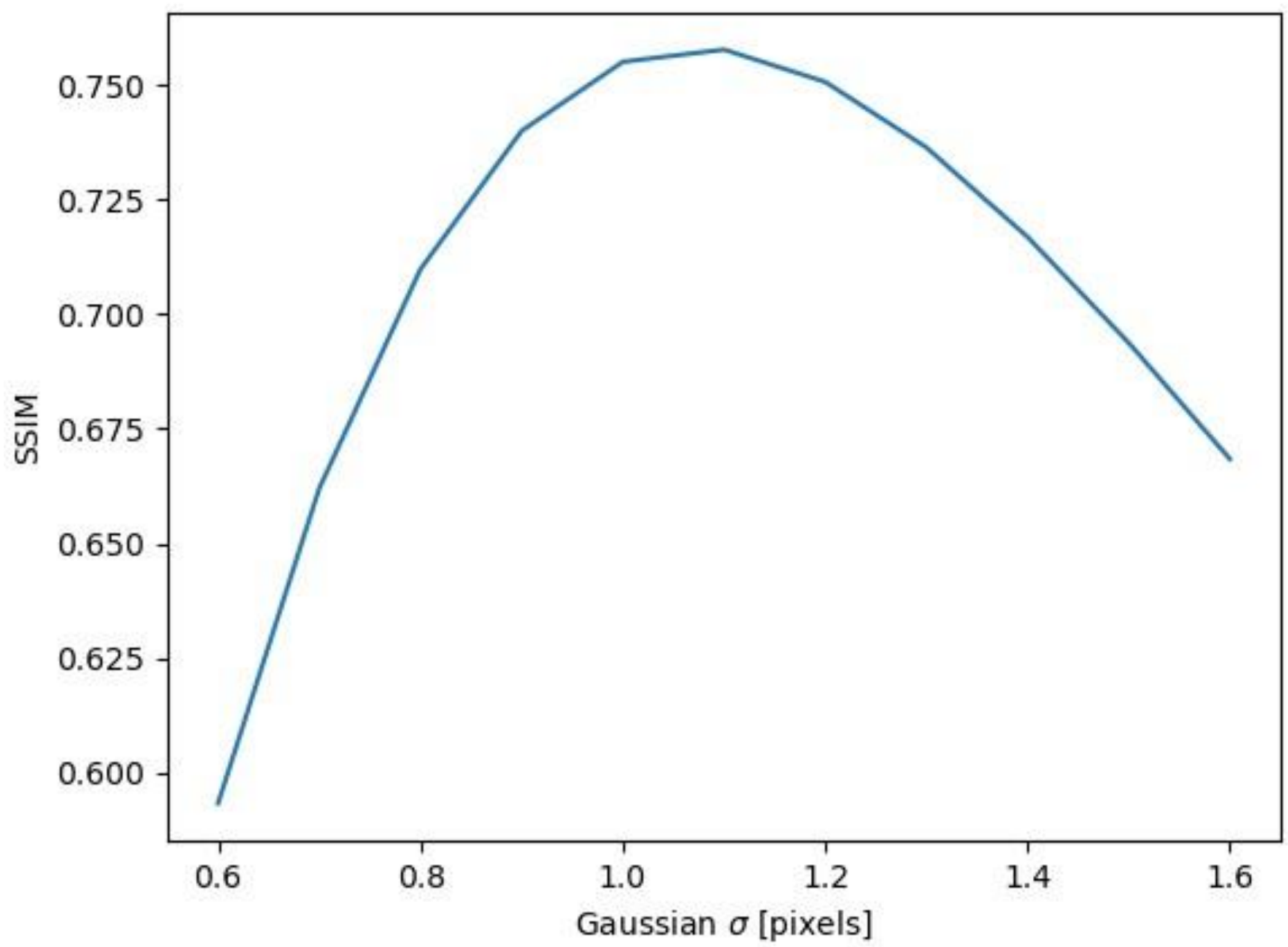

**Figure S2**: **Dependence of structural similarity index measure (SSIM) on Gaussian σ.** The SSIM has been computed for 2000 images with various noise level. The plotted dependence enables choice of the optimal standard deviation σ for the Gaussian denoising approach.

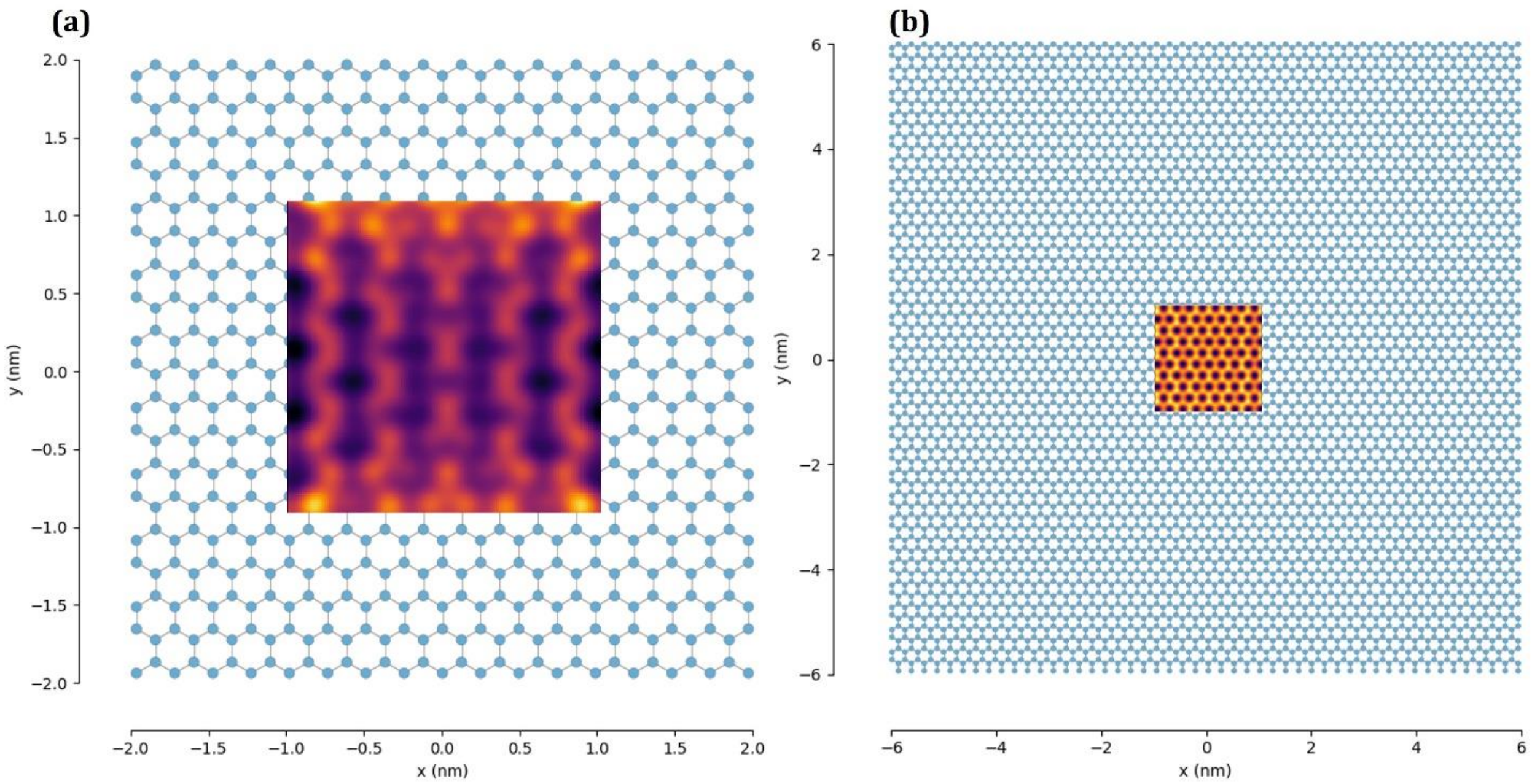

**Figure S3**: **Illustration of the method to induce quasiparticle interference (QPI) patterns in the simulated STM image**. To introduce QPI patterns, it suffices to consider graphene nanosheet which are only slightly larger than the area considered to produce the STM image (located at the center of the graphene nanosheet). **(a)** The nanosheet is 2 nm larger than the STM image (4 nm vs. 2 nm). **(b)** The nanosheet is 10 nm larger than the STM image (12 nm vs. 2 nm). The other parameters are the same. One can see that QPI patterns are prominent in panel (a), whereas almost no QPI pattern is visible in panel (b).

## References

[1] A. Mahmood, P. Mallet, and J.-Y. Veuillen, *Quasiparticle Scattering off Phase Boundaries in Epitaxial Graphene*, Nanotechnology **23**, 055706 (2012).

[2] J. Tesch, P. Leicht, F. Blumenschein, L. Gragnaniello, A. Bergvall, T. Löfwander, and M. Fonin, *Impurity Scattering and Size Quantization Effects in a Single Graphene Nanoflake*, Phys Rev B **95**, 75429 (2017).